# Electromagnetic structure of hadrons

**Alexander G. Kyriakos**

*Saint-Petersburg State Institute of Technology, St.Petersburg, Russia*

Present address:
*Athens, Greece*
*e-mail:* lelekous@otenet.gr

## Abstract

In the previous papers [1,2,3], based on Dirac's equation, we have considered the electromagnetic structure of the leptons. In the present paper, using the Yang-Mills equation, we will analyse the electromagnetic structure of the hadrons.

## Contents

# 1. Introduction

As it is known the modern Standard Model Theory of elementary particles is described by the Yang-Mills equation [4].

## 1.1 The Yang-Mills equation and QCD

The Standard Model Theory contains the electroweak theory, which is based on the SU(2)xU1 symmetry group, and the strong interaction theory (SU(3) symmetry group). Both theories are built on the base of the Yang-Mills equation in the same way.

Let’s introduce briefly the QCD theory [4,5,6]. As it is known, the QCD theory and the QED theory forms are alike. Particularly, the Lagrangian and the equation of the QCD have the same form as the Lagrangian and the equation of the QED [4,5].

Since it is important for our theory understanding, we consider at first the theory structure of QED.

In the usual notations [5] the Lagrangian of QED is represented by the sum:

$$L_{QED} = \bar{\psi}\left[\left(i\partial_\mu + eA_\mu\right)\gamma_\mu - m\right]\psi - \frac{1}{4}F_{\mu\nu}F_{\nu\mu}\,, \tag{1.1}$$

where $\partial_\mu = \partial / \partial\, x_\mu$ is the partial derivative with respect to 4-co-ordinate $x_\mu$, $F_{\mu\nu} = \partial_\mu A_\nu - \partial_\nu A_\mu$ is the operator of electromagnetic field strength, $-e,\ m$ are the electrical charge and the mass of the electron, respectively; $\gamma_\mu$ are the Dirac's matrices, $\bar{\psi}$ is the hermitian wave function, $A_\mu$ is the potential of the electromagneticl field (the summation is always taken with respect to same indeces). The first and the third summands of the Lagrangian describe the free motion of electron, the last summand describes the same of photons, and the term $\bar{\psi}A_\mu\psi$ describes its interaction. Using the covariant derivative

$$D_\mu = \partial_\mu - ieA_\mu\,, \tag{1.2}$$

we can write the Lagrangian in the form:

$$L_{QED} = \bar{\psi}\left[D_\mu\gamma_\mu - m\right]\psi - \frac{1}{4}F_{\mu\nu}F_{\nu\mu}\,, \tag{1.3}$$

The physical means of the present theory will be understood better, if we use the our notations [1]. In this notations we have for QED Lagrangian:

$$L_{QED} = L_D + L_{\text{int}} + L_M\,, \tag{1.4}$$

where

$$L_D = \psi^+\left(\hat{\alpha}_0\hat{\varepsilon} + c\hat{\vec{\alpha}}\ \hat{\vec{p}} + \hat{\beta}\ m_e c^2\right)\psi, \tag{1.5}$$

is the free electron Lagrangian.

The Lagrangian:

$$L_{\text{int}} = -\psi^+\left(\hat{\alpha}_0\varepsilon_{ex} + c\hat{\vec{\alpha}}\ \vec{p}_{ex}\right)\psi\,, \tag{1.6}$$

is the interaction Lagrangian, and

$$L_M = \frac{1}{8\pi}\left(\vec{E}^2 - \vec{H}^2\right), \qquad (1.7)$$

is the Maxwell theory Lagrangian.

Thus, the full Lagrangian of QED theory is:

$$L_{QED} = \psi^+\left[\hat{\alpha}_0(\hat{\varepsilon} - \varepsilon_{ex}) + c\hat{\vec{\alpha}}\left(\hat{\vec{p}} - \vec{p}_{ex}\right) + \hat{\beta}\, m_e c^2\right]\psi + \frac{1}{8\pi}\left(\vec{E}^2 - \vec{H}^2\right), \qquad (1.8)$$

where $\hat{\alpha}_0, \hat{\vec{\alpha}}, \hat{\beta}$ - are Dirac's matrices, $\hat{\varepsilon} = i\hbar\frac{\partial}{\partial t}$, $\hat{\vec{p}} = -i\hbar\vec{\nabla}$ - the operators of energy and momentum, $\psi^+$ is the Hermitian wave function, $\varepsilon_{ex} = e\varphi$, $\vec{p}_{ex} = \frac{e}{c}\vec{A}$ are the external electron energy and momentum , $c$ is the light velocity, $m_e$ is the electron mass, $\vec{E}$, $\vec{H}$ are the electric and magnetic field, respectively, and $\left(\varphi, \vec{A}\right)$ is 4-potential of external field. The matrices $\hat{\alpha}_0, \hat{\vec{\alpha}}$, $\hat{\beta}$ are the following Dirac’s matrices:

$$\hat{\alpha}_0 = \begin{pmatrix} 1 & 0 & 0 & 0 \\ 0 & 1 & 0 & 0 \\ 0 & 0 & 1 & 0 \\ 0 & 0 & 0 & 1 \end{pmatrix}, \quad \hat{\alpha}_1 = \begin{pmatrix} 0 & 0 & 0 & 1 \\ 0 & 0 & 1 & 0 \\ 0 & 1 & 0 & 0 \\ 1 & 0 & 0 & 0 \end{pmatrix}, \quad \hat{\alpha}_4 \equiv \hat{\beta} = \begin{pmatrix} 1 & 0 & 0 & 0 \\ 0 & 1 & 0 & 0 \\ 0 & 0 & -1 & 0 \\ 0 & 0 & 0 & -1 \end{pmatrix}, \qquad (1.9)$$

$$\hat{\alpha}_2 = \begin{pmatrix} 0 & 0 & 0 & -i \\ 0 & 0 & i & 0 \\ 0 & -i & 0 & 0 \\ i & 0 & 0 & 0 \end{pmatrix}, \quad \hat{\alpha}_3 = \begin{pmatrix} 0 & 0 & 1 & 0 \\ 0 & 0 & 0 & -1 \\ 1 & 0 & 0 & 0 \\ 0 & -1 & 0 & 0 \end{pmatrix}$$

In Quantum Chromodynamics (QCD) we have quarks instead of electrons and gluons instead of photons, between which there are the strong interactions instead of electromagnetic interactions. The strong interactions among quarks and gluons are described by the non-abelian gauge theory, based on the gauge group $SU(3)_C$, instead of the abelian gauge theory of the EM theory, based on the gauge group $U(1)$. Each quark sort (flavour) corresponds to a colour (strong interaction charge) triplet in the fundamental representation of $SU(3)$ and the gauge fields needed to maintain the gauge symmetry, the gluons, are in the adjoint representation of dimension 8. Gauge invariance ensures that gluons are massless whose spin is equal to 1, as the photon. Since the gluons masses are equal to zero, the quark interaction radius is equal to infinity, but the sizes of the quark system (particles) are very small (in order $10^{-13}$ cm). The explanation of this fact is not known (the «confinement» problem).

The QCD Lagrangian may by written in ordinary notations [4,5,6] as

$$L_{QCD} = \overline{\psi}_i\left[D_\mu\gamma_\mu - m\right]\psi_i - \frac{1}{4}F_{\mu\nu}F_{\nu\mu}, \qquad (1.10)$$

where $$F_{\mu\nu} = F^a_{\mu\nu}\frac{\lambda_a}{2}, \qquad (1.11)$$

stands for the gluon field tensor, $\psi_i$ are the quark fields and the covariant derivative is defined by

$$D_\mu = \partial_\mu - igA_\mu, \qquad (1.12)$$

The strong coupling is represented by $g$ and indices are summed over $a = 1,...,8$ and over $i = 1,2,3$. Finally, $\lambda_a/2$ and $f_{abc}$ are the $SU(3)$ group generators and structure constants, respectively, which are related to the commutator:

$$[\lambda_a, \lambda_b] = 2if_{abc}\lambda^c, \qquad (1.13)$$

For the gluons the field strength has the view:

$$F_{\mu\nu} = \partial_\mu A_\nu - \partial_\nu A_\mu - ig[A_\mu A_\nu - A_\nu A_\mu], \qquad (1.14)$$

or in the vector form:

$$\vec{F}_{\mu\nu} = \partial_\mu \vec{A}_\nu - \partial_\nu \vec{A}_\mu + g\vec{A}_\mu \times \vec{A}_\nu, \qquad (1.15)$$

The photon potentials $A_\mu$ are the numbers and in the QED case the above commutator is equal to zero. In the case of the non-abelian gauge fields (gluons) $A_\mu$ are the matrices, and the above comutator is not equal to zero. Thanks to this commutator, the non-linear interaction of the gluons appears.

The matrices $\lambda_a$ are named Gell-Mann matrices and have the view:

$$\lambda_1 = \begin{pmatrix} 0 & 1 & 0 \\ 1 & 0 & 0 \\ 0 & 0 & 0 \end{pmatrix}, \quad \lambda_2 = \begin{pmatrix} 0 & -i & 0 \\ i & 0 & 0 \\ 0 & 0 & 0 \end{pmatrix}, \quad \lambda_3 = \begin{pmatrix} 1 & 0 & 0 \\ 0 & -1 & 0 \\ 0 & 0 & 0 \end{pmatrix},$$

$$\lambda_4 = \begin{pmatrix} 0 & 0 & 1 \\ 0 & 0 & 0 \\ 1 & 0 & 0 \end{pmatrix}, \quad \lambda_5 = \begin{pmatrix} 0 & 0 & -i \\ 0 & 0 & 0 \\ i & 0 & 0 \end{pmatrix}, \quad \lambda_6 = \begin{pmatrix} 0 & 0 & 0 \\ 0 & 0 & 1 \\ 0 & 1 & 0 \end{pmatrix}, \qquad (1.16)$$

$$\lambda_7 = \begin{pmatrix} 0 & 0 & 0 \\ 0 & 0 & -i \\ 0 & i & 0 \end{pmatrix}, \quad \lambda_8 = \begin{pmatrix} 1 & 0 & 0 \\ 0 & 1 & 0 \\ 0 & 0 & -2 \end{pmatrix}.$$

If we consider the quark as real fermion, the contradiction with Pauli principle appears. To eliminate this contradiction some inner degree of freedom is introduced, named *colour (*or *colour charge)*.

According to modern theory in strong interactions the colour charges play the same role as the electrical charges in the electromagnetic interactions. The interaction between electrons takes place through the photon exchange. In this case gluons play the role of photons, which are also electrically neutral vector particles. The basic difference between gluons and photons is that the photon is one, but gluons are 8 and have colour charges.

Thanks to these colour charges the gluons interact strongly with one another and can radiate one another.

## 1.2 The investigation object choice

Apparently the simplest of the hadrons must be the scope of our investigation. There is a reason to believe that the simplest particles are particles with the lower mass of the mass spectrum of each particle family. In the case of the baryons the simplest particles are nucleons - proton (antiproton) and neutron (antineutron); among the mesons these are the pi mesons. We will name these particles the *basic particles*.

As it is known, the basic particles consist only of two quarks $u$ and $d$:

$$\pi^+ = u\bar{d}, \quad \pi^- = d\bar{u}, \quad \pi^0 = \frac{1}{\sqrt{2}}\left(u\bar{u} - d\bar{d}\right), \quad p = uud, \quad n = ddu$$

which have the following charges:

$$q_u = \frac{2}{3}e, \quad q_d = -\frac{1}{3}e$$

The quark masses don't depend on their colour, but depend on the flavour. The $u$ and $d$ quarks have aproximetly the following masses:

$$m_u \approx 4\ MeV, \quad m_d \approx 7\ MeV$$

## 1.3 Problem statement

The problem of the present paper is to show the electromagnetic structure of the hadrons. As it is followed from the Standard Model Theory (SMT), the quark family is analogue to the lepton family.

Based on this analogy we suppose that:

1) **The electromagnetic structure of the hadrons is similar to the lepton structure;**
2) **the above hadron Lagrangian (and equations) is composed from three Lagrangians (equations) of the lepton type, i.e. by three Dirac's equation Lagrangians (three Dirac's equations).**

*In other words, we must show that the lepton Lagrangian (also lepton equation) is the «one quark» Lagrangian (equation).*

# 2. “One quark” theory

## 2.1 Inner and external fields

The comparison between the QCD and the QED Lagrangians, that we made above, is not entirely right. The QED Lagrangian is not the free lepton Lagrangian, but the Lagrangian with interaction between leptons and photons. On the other hand, the above Lagrangian of hadrons is indeed the Lagrangian of the free hadron.

*This distinction is bounded with the distinction among inner and external fields of the particles. The external field, used in*

*the QED for the description of the interaction among the electron and other charge particles, is in the case of hadron the inner field, describing the quark-quark interaction.*

Thus, it follows from above that the “one quark” Lagrangian and equation must be the QED Lagrangian (equation) for external field, but inside the hadron this “external field” describes the quark-quark interaction.

## 2.2 The “one quark” equation

For the solution of the present problem we use here some results, which were represented in detail in the paper [1]. Particularly, to understand the “one quark” equation form, we consider the electromagnetic and quantum forms of electron-positron pair production theory.

Let's consider the particle-antiparticle production conditions. One $\gamma$-quantum cannot turn spontaneously into the electron-positron pair, although it interacts with the electron-positron vacuum. For the pair production, at first, the following mass correlation is necessary: $\varepsilon_p \geq 2m_e c^2$ (where $\varepsilon_p$ is the photon energy, $m_e$ - the electron mass and $c$ - the light velocity). At second, the presence of the other particle, having the electromagnetic field, is needed. It can be some other $\gamma$-quantum, the electron $e^-, e^+$, an atom nucleus $Ze$ etc. For example, we have the typical reaction:

$$\gamma + Ze \to Ze + e^+ + e^- , \qquad (2.1)$$

which means that, while moving through the particle field (as in medium with high refraction index) the photon (or maybe virtual photon) takes some transformation, which corresponds to the pair production. Considering the fact that Pauli's matrices describe the vector rotations and also taking in account the optical-mechanical analogy analysis, we can assume that the above transformation is a field distortion. From this follows the **distortion hypothesis**:

***By the fulfilment of the pair production conditions the distortion of the electromagnetic field of the photon can take place; as a result photon is able to move along the closed trajectory, making some stable construction named elementary particle.***

*About the electron, as the simplest particle, we can suppose that the photon trajectory is circular.*

Consider the linear photon, moving along $-y$-axis (of course, we can also use any other direction [2]). In a general case it has the two possible polarisations and contains the field vectors $E_x, E_z, H_x, H_z$ $(E_y = H_y = 0)$. Such photon can form the ring only on the $(x,o,y)$ or the $(y,o,z)$ plains.

The bispinor form of free lepton Dirac's equations can be written as one equation [7]:

$$\hat{\varepsilon}\psi + c\hat{\vec{\alpha}}\,\hat{\vec{p}} + \hat{\beta}\, m_e c^2 \psi = 0, \qquad (2.2)$$

(the notations see above).

Put the following semi-photon bispinor:

$$\psi = \begin{pmatrix} E_z \\ E_x \\ iH_z \\ iH_x \end{pmatrix} \qquad (2.3)$$

Using (2.3), we can write the equation of the electromagnetic wave moved along any axis in form:

$$\left(\hat{\varepsilon}^{\,2} - c^2 \hat{\vec{p}}^{\,2}\right)\psi = 0, \qquad (2.4)$$

The equation (2.4) can also be written in the following form:

$$\left[\left(\hat{\alpha}_o \hat{\varepsilon}\right)^2 - c^2\left(\hat{\vec{\alpha}}\,\hat{\vec{p}}\right)^2\right]\psi = 0, \qquad (2.5)$$

where $\hat{\alpha}_o = \hat{1}$ is the unit matrix. In fact, taking into account that

$$\left(\hat{\alpha}_o \hat{\varepsilon}\right)^2 = \hat{\varepsilon}^{\,2}, \quad \left(\hat{\vec{\alpha}}\,\hat{\vec{p}}\right)^2 = \hat{\vec{p}}^{\,2},$$

we see that equations (2.4) and (2.5) are equivalent.

Factorising (2.5) and multiplying it from left on Hermithian-conjugate function $\psi^+$ we get:

$$\psi^+\left(\hat{\alpha}_o \hat{\varepsilon} - c\hat{\vec{\alpha}}\,\hat{\vec{p}}\right)\left(\hat{\alpha}_o \hat{\varepsilon} + c\hat{\vec{\alpha}}\,\hat{\vec{p}}\right)\psi = 0, \qquad (2.6)$$

The equation (2.6) may be disintegrated on two equations:

$$\psi^+\left(\hat{\alpha}_o \hat{\varepsilon} - c\hat{\vec{\alpha}}\,\hat{\vec{p}}\right) = 0, \qquad (2.7)$$

$$\left(\hat{\alpha}_o \hat{\varepsilon} + c\hat{\vec{\alpha}}\,\hat{\vec{p}}\right)\psi = 0, \qquad (2.8)$$

It is not difficult to show (using (2.2)) that the equations (2.7) and (2.8) are Maxwell's equations without current and, at the same time, are Dirac's electron-positron equations without mass.

*In accordance with our assumption, the reason for current appearance must be the electromagnetic wave motion along a curvilinear trajectory.* We will show the appearance of the current, using the general methods of the distortion field investigation [8], but the same result can be obtained simplier in the vector form (see [1] Appendix 1, chapter A1.2.). The question is about the tangent space introduction at every point of the curvilinear space, in which the orthogonal axis system moves. This corresponds to the fact, that the wave motion along a circular trajectory is accompanied by the motion of the rectangular basis, built on vectors ($\vec{E}, \vec{S}, \vec{H}$), where $\vec{S}$ is the Poynting vector.

For the generalisation of Dirac's equation in Riemann’s geometry it is necessary [8] to replace the usual derivative $\partial_\mu \equiv \partial / \partial x_\mu$ (where $x_\mu$ are the co-ordinates in the 4-space) with the covariant derivative: $D_\mu = \partial_\mu + \Gamma_\mu$ ($\mu = 0, 1, 2, 3$ are the summing indexes), where $\Gamma_\mu$ is the analogue of Christoffel's symbols in the case of the spinors theory. When a spinor moves along the beeline, all $\Gamma_\mu = 0$, and we have a usual derivative. But if a spinor moves along the curvilinear trajectory, then not all $\Gamma_\mu$ are equal to zero and a supplementary term appears. Typically, the last one is not the derivative, but it is equal to the product of the spinor itself with some coefficient $\Gamma_\mu$. Thus we can assume that the supplementary term is a longitudinal field, i.e. a current. So from (2.7-2.8) we obtain:

$$\alpha^\mu D_\mu \psi = \alpha^\mu (\partial_\mu + \Gamma_\mu)\ \psi = 0\,,$$

According to the general theory [8] the increment in spinor $\Gamma_\mu$ has the form of the energy-momentum 4-vector. It is logical (see also [1], Appendix 1, chapter A1.2.) to identify $\Gamma_\mu$ with 4-vector of energy-momentum of the electron's own field:

$$\Gamma_\mu = \{\varepsilon_s, c\vec{p}_s\}\,, \tag{2.9}$$

Then equations (2.7) and (2.8) in the curvilinear space will have the view:

$$\psi^+[\ \left(\hat{\alpha}_o\hat{\varepsilon} - c\hat{\vec{\alpha}}\ \hat{\vec{p}}\right) - \left(\hat{\alpha}_o\varepsilon_s - c\hat{\vec{\alpha}}\ \vec{p}_s\right)\ ] = 0\,, \tag{2.10}$$

$$[\ \left(\hat{\alpha}_o\hat{\varepsilon} + c\hat{\vec{\alpha}}\ \hat{\vec{p}}\right) + \left(\hat{\alpha}_o\varepsilon_s + c\hat{\vec{\alpha}}\ \vec{p}_s\right)\ ]\ \psi = 0\,, \tag{2.11}$$

According to the energy conservation law we can write:

$$\hat{\alpha}_o\varepsilon_s \pm c\hat{\vec{\alpha}}\ \vec{p}_s = \mp\hat{\beta}\ m_e c^2\,, \tag{2.12}$$

Substituting (2.12) in (2.10) and (2.11) we will arrive at the usual kind of Dirac's equation with the mass:

$$\psi^+[\ \left(\hat{\alpha}_o\hat{\varepsilon} - c\hat{\vec{\alpha}}\ \hat{\vec{p}}\right) - \hat{\beta}\ m_e c^2\ \ ] = 0\,, \tag{2.13}$$

$$[\ \left(\hat{\alpha}_o\hat{\varepsilon} + c\hat{\vec{\alpha}}\ \hat{\vec{p}}\right) + \hat{\beta}\ m_e c^2\ \ ]\ \psi = 0\,, \tag{2.14}$$

Figure 1 illustrates the above description of the process of electron-positron pair generation:

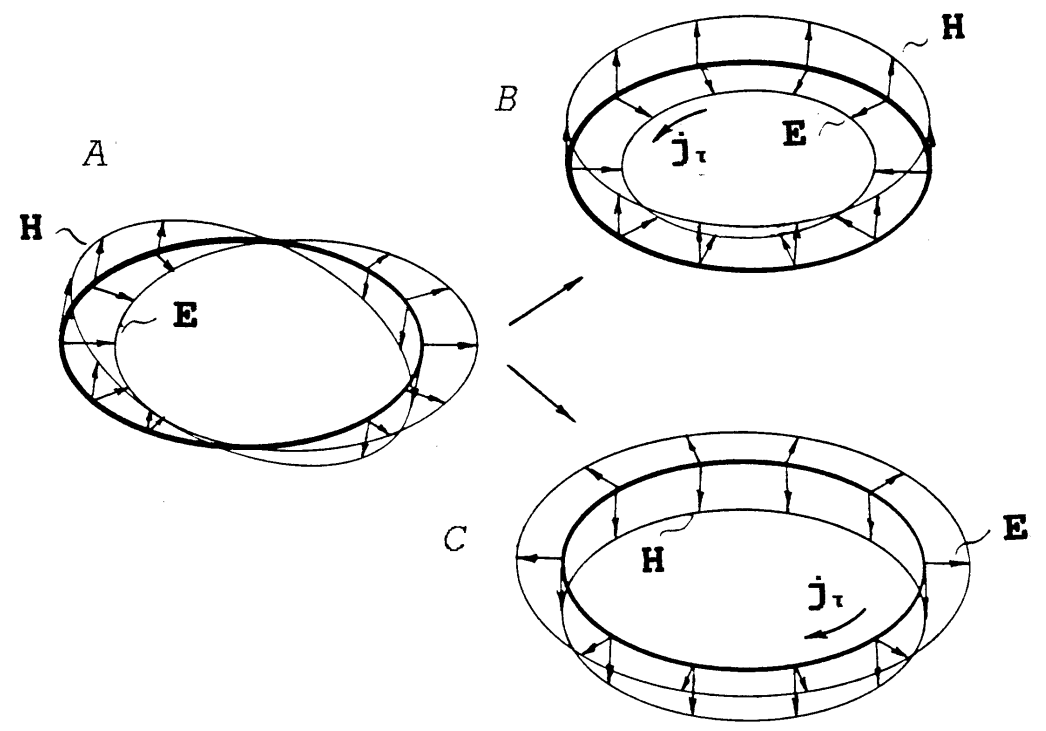

**Fig.1**

*It is not difficult to see that the pair production process corresponds to the photon division on two circular semi-periods (semi-photons). According to our supposition the "one quark" is the ring or knot (loop) of the same type.*

**2.3. Electromagnetic form of “one quark” equation**

Taking into account that $\psi = \psi(y)$, from (2.13) using (2.14) we obtain:

$$\begin{cases} rot\ \vec{E} + \frac{1}{c}\frac{\partial\ \vec{H}}{\partial\ t} = i\frac{\omega}{c}\vec{H}, \\ rot\ \vec{H} - \frac{1}{c}\frac{\partial\ \vec{E}}{\partial\ t} = -i\frac{\omega}{c}\vec{E}, \end{cases}, \qquad (2.15)$$

where $\omega = \frac{m_e c^2}{\hbar}$. The equations (2.15) are the Maxwell equations with current [9]. It is interesting that along with the electrical current the magnetic current also exists here. This current is equal to zero by Maxwell's theory, but its existence by Dirac doesn't contradict to the quantum theory. (As we have showed [3] the magnetic current appearance relates to the initial photon circular polarisation and integrally this current is equal to zero).

**2.4. The “one quark” linear equation Lagrangian**

As it is known [9], the Lagrangian of the free field Maxwell's theory is:

$$L_M = \frac{1}{8\pi}\left(\vec{E}^2 - \vec{H}^2\right) \qquad (2.16)$$

The following expression can be taken as Lagrangian of Dirac's theory [7]:

$$L_D = \psi^+\left(\hat{\alpha}_0\hat{\varepsilon} + c\hat{\vec{\alpha}}\ \hat{\vec{p}} + \hat{\beta}\ m_e c^2\right)\psi, \qquad (2.17)$$

For the wave moving along the $y$-axis the equation (2.17) can be written:

$$L_D = \frac{1}{c}\psi^+\frac{\partial\ \psi}{\partial\ t} - \psi^+\hat{\alpha}_y\frac{\partial\ \psi}{\partial\ y} - i\frac{m_e c}{\hbar}\psi^+\hat{\beta}\ \psi, \qquad (2.18)$$

Transferring each term of (2.18) in electrodynamics' form (see [1]) we obtain for the semi-photon or “one quark” equation the following Lagrangian:

$$L_D = \psi^+\left(\hat{\alpha}_0\hat{\varepsilon} + c\hat{\vec{\alpha}}\ \hat{\vec{p}}\right)\psi - i\frac{\omega_s}{8\pi}\left(\vec{E}^2 - \vec{H}^2\right), \qquad (2.19')$$

or in ordinary form:

$$L_D = \psi^+\alpha_\mu\partial_\mu\psi - i\frac{\omega_s}{8\pi}\left(\vec{E}^2 - \vec{H}^2\right), \qquad (2.19'')$$

where $\omega_s = \frac{2m_e c^2}{\hbar}$ (note that we must differ the complex conjugate field vectors $\vec{E}^*$, $\vec{H}^*$ and $\vec{E}$, $\vec{H}$).

As we have supposed, the Lagrangian (2.19′′) is the Lagrangian of “one quark”, and it is actually similar to (1.10).

*We can say that the Lagrangian (1.10) is constructed from Lagrangians of three interacting quarks. Therefore, to obtain (1.10) we must sum three Lagrangians (2.19′′) and "turn on" the interactions among quarks (i.e. to go over from simple derivative to the covariant derivative).*

Now we will show that the photon interaction term $\frac{1}{4}F_{\mu\nu}F_{\nu\mu}$ can be represented as non-linear expression the same type as (1.15), which describes the photon-photon (i.e. gluon-gluon) interactions.

## 2.5. Lagrangian of non-linear "one quark" equation

Using (2.10) we can write (see [1]) the Lagrangian of "one quark" equation in the form:

$$L_N = \psi^+\left(\hat{\varepsilon} - c\hat{\vec{\alpha}}\,\hat{\vec{p}}\right)\psi + \psi^+\left(\varepsilon_s - c\hat{\vec{\alpha}}\,\vec{p}_s\right)\psi, \tag{2.20}$$

Let's show that **the expression (2.20) represents the common form of Lagrangian of non-linear "one quark" (or "one ring", or "one knot", or electron Dirac's) equation.**

Taking into account that the free electron Dirac's equation solution is the plane wave:

$$\psi = \psi_0\, e^{i(\omega t - ky)}, \tag{2.21}$$

we can represent (2.20) in the approximate quantum form:

$$L_N = \psi^+\hat{\alpha}_\mu\partial_\mu\psi + \frac{\Delta\tau_s}{8\pi}\left[\left(\psi^+\psi\right)^2 - \left(\psi^+\hat{\vec{\alpha}}\psi\right)^2\right], \tag{2.22}$$

where $\Delta\tau_s$ is the "volume" of the particle.

Transform (2.20) into electrodynamics form. For energy and momentum in the electromagnetic form we have [1]:

$$\varepsilon_s = \int_{\Delta\tau} U\, d\tau = \frac{1}{8\pi}\int_{\Delta\tau}\left(\vec{E}^2 + \vec{H}^2\right)d\tau, \tag{2.23}$$

$$\vec{p}_s = \int_{\Delta\tau}\vec{g}\, d\tau = \frac{1}{c^2}\int_{\Delta\tau}\vec{S}\, d\tau = \frac{1}{4\pi}\int_{\Delta\tau}\left[\vec{E}\times\vec{H}\right]d\tau, \tag{2.24}$$

Normalising $\psi$-function by relationship:

$$L'_N = \frac{1}{8\pi\, m_e c^2}L_N, \tag{2.25}$$

and using the equations (2.23) and (2.24), we find:

$$L'_N = \psi^+\hat{\alpha}_\mu\partial_\mu\psi + 2\Delta\tau\left(U^2 - c^2\vec{g}^2\right), \tag{2.26}$$

It is not difficult to transform the second summand, using the known electrodynamics' transformation (that is *quantum form of the known Fierz's correlation)*:

$$(8\pi)^2\left(U^2 - c^2\vec{g}^2\right) = \left(\vec{E}^2 + \vec{H}^2\right)^2 - 4\left(\vec{E}\times\vec{H}\right)^2 = \left(\vec{E}^2 - \vec{H}^2\right)^2 + 4\left(\vec{E}\cdot\vec{H}\right)^2, \tag{2.27}$$

So we have:

$$L'_N = \psi^+\hat{\alpha}_\mu\partial_\mu\psi + 2\Delta\tau\left[\left(\vec{E}^2 + \vec{H}^2\right)^2 - 4\left(\vec{E}\times\vec{H}\right)^2\right], \tag{2.28}$$

or

$$L_N^{'} = \psi^+ \hat{\alpha}_\mu \partial_\mu \psi + 2\Delta\tau \left[ \left( \vec{E}^2 - \vec{H}^2 \right)^2 + 4 \left( \vec{E} \cdot \vec{H} \right)^2 \right], \qquad (2.28')$$

We can say that this Lagrangian is the "one quark" non-linear equation Lagrangian.

Actually, as we have shown in the paper [1], the Lagrangian (2.28) is similar to the Lagrangian of the photon-photon interaction [10]:

$$L_{p-p} = \frac{1}{8\pi} \left( \vec{E}^2 - \vec{H}^2 \right) + b \left[ \left( \vec{E}^2 - \vec{H}^2 \right)^2 + 7 \left( \vec{E} \cdot \vec{H} \right)^2 \right] + \ldots, \qquad (2.29)$$

where $b$ is constant.

Using (2.28) we can build the "three quark" Lagrangian of the same type as (1.10) with (1.15).

# 3. Quantum and electromagnetic forms of "three quark" equations

## 3.1 Electromagnetic forms of quantum equations

It has been known for a long time [10,11] that the quantum equations can formally be represented as the Maxwell equation system. For example, according to [11] the spinor Dirac's equation system

$$\begin{cases} \left[ \left( \hat{\sigma}_o \hat{\varepsilon} - c \hat{\vec{\sigma}} \; \hat{\vec{p}} \right) - mc^2 \right] \varphi = 0 \\ \left[ \left( \hat{\sigma}_o \hat{\varepsilon} + c \hat{\vec{\sigma}} \; \hat{\vec{p}} \right) + mc^2 \right] \chi = 0 \end{cases}, \qquad (3.1)$$

becomes the full Maxwell equation system, if we use instead of $\varphi$ and $\chi$ 2x1-matrix wave functions, the 3x1-matrix electromagnetic field function:

$$\varphi = \left( \vec{E} \right), \quad \chi = \left( i\vec{H} \right), \qquad (3.2)$$

and instead of 2x2-spinor Pauli's matrices $\hat{\sigma}$ we use the following 3x3-matrices:

$$\hat{S}_1 = \begin{pmatrix} 0 & 0 & 0 \\ 0 & 0 & -i \\ 0 & i & 0 \end{pmatrix}, \quad \hat{S}_2 = \begin{pmatrix} 0 & 0 & i \\ 0 & 0 & 0 \\ i & 0 & 0 \end{pmatrix}, \quad \hat{S}_3 = \begin{pmatrix} 0 & -i & 0 \\ i & 0 & 0 \\ 0 & 0 & 0 \end{pmatrix}, \qquad (3.3)$$

The zero and unit 3x3-matrices are also useful:

$$\hat{S}_0 = \begin{pmatrix} 1 & 0 & 0 \\ 0 & 1 & 0 \\ 0 & 0 & 1 \end{pmatrix}, \quad \hat{0} = \begin{pmatrix} 0 & 0 & 0 \\ 0 & 0 & 0 \\ 0 & 0 & 0 \end{pmatrix}, \qquad (3.4)$$

As it is known [4], these matrices are the rotation generators of SU(2) group and coincide with some matrices of the chromodynamics.

We will describe two equations (3.1) as one equation:
for particle

$$[\ \left({}^6\hat{\alpha}_o\hat{\varepsilon} - c\,{}^6\hat{\vec{\alpha}}\ \hat{\vec{p}}\right) - {}^6\hat{\beta}\ mc^2\ ]\ \psi = 0\,, \qquad (3.5)$$

and for antiparticle

$$\psi^+[\ \left({}^6\hat{\alpha}_o\hat{\varepsilon} + c\,{}^6\hat{\vec{\alpha}}\ \hat{\vec{p}}\right) + {}^6\hat{\beta}\ mc^2\ ] = 0\,, \qquad (3.6)$$

where upper left index "6" means that these matrices are the 6x6-matrices of the following type:

$${}^6\hat{\vec{\alpha}} = {}^6\hat{\alpha}_x\vec{i} + {}^6\hat{\alpha}_y\vec{j} + {}^6\vec{\alpha}_z\vec{k} = \begin{pmatrix} \hat{0} & \hat{\vec{S}} \\ \hat{\vec{S}} & \hat{0} \end{pmatrix},\quad {}^6\hat{\alpha}_0 = \begin{pmatrix} \hat{S}_0 & \hat{0} \\ \hat{0} & \hat{S}_0 \end{pmatrix},\quad {}^6\hat{\alpha}_4 \equiv {}^6\hat{\beta} = \begin{pmatrix} \hat{S}_0 & \hat{0} \\ \hat{0} & -\hat{S}_0 \end{pmatrix}, \qquad (3.7)$$

The wave function has the 6x1-matrix form:

$${}^6\psi = \begin{pmatrix} \vec{E} \\ i\vec{H} \end{pmatrix}, \qquad (3.8)$$

We name the equations (3.5)-(3.6) three-knot equations and 6x6-matrices - three-knot matrices.

The above matrices give the right expressions of the bilinear form of the theory (i.e. the basic values of Maxwell's theory), as it is not difficult to test:

for the energy:

$${}^6\psi^+\ {}^6\hat{\alpha}_0\,{}^6\psi = \vec{E}^2 + \vec{H}^2 = 8\pi\ U\,, \qquad (3.9)$$

for the momentum projections of the electromagnetic field:

$$\begin{aligned} {}^6\psi^+\ {}^6\hat{\alpha}_1\,{}^6\psi &= 2(E_yH_z - E_zH_y)\,, \\ {}^6\psi^+\ {}^6\hat{\alpha}_2\,{}^6\psi &= 2(E_zH_x - E_xH_z)\,, \\ {}^6\psi^+\ {}^6\hat{\alpha}_3\,{}^6\psi &= 2(E_xH_y - E_yH_x)\,, \end{aligned} \qquad (3.10)$$

or for the Poynting vector:

$$\vec{S}_P = \frac{1}{4\pi}\begin{vmatrix} \vec{i} & \vec{j} & \vec{k} \\ E_x & E_y & E_z \\ H_x & H_y & H_z \end{vmatrix} = \qquad ,\ (3.11)$$

$$= \frac{1}{4\pi}\left\{\vec{i}(E_yH_z - E_zH_y) - \vec{j}(E_xH_z - E_zH_x) + \vec{k}(E_xH_y - E_yH_x)\right\} == \frac{1}{8\pi}\,{}^6\psi^+\,{}^6\vec{\alpha}\ {}^6\psi$$

and for 1st scalar of electromagnetic field:

$${}^6\psi^+\ {}^6\hat{\alpha}_4\,{}^6\psi = \vec{E}^2 - \vec{H}^2 = 8\pi\ I_1\,, \qquad (3.12)$$

### 3.2 Rings or knots?

We consider the proton as three "one quark" construction, i.e. as three bounded lepton constructions, each of which is a ring.

According to above suppositions the proton model can have the following forms (Fig.2):

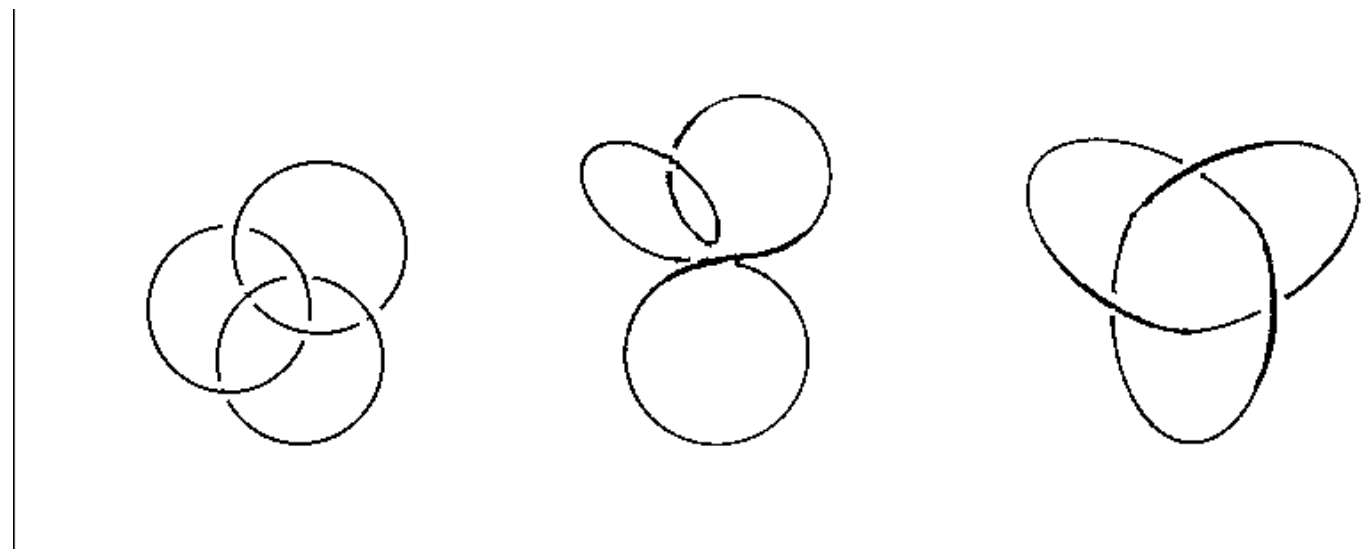

Fig.2

A first question appears: does the proton consist of rings or of knots?

The following argument exists about the proton consisting of the three engaged knots: if the proton consisted of rings, the proton charge would be equal to $3e$. Therefore, proton has the third scheme of the figure 2. (On the other hand the asymptotic freedom corresponds better to the ring structure of the barions).

### 3.2 "Three quarks" equation without interaction

From the above follows that the proton equation contains three "one quark" equations, i.e. three electron equations or three pair of the scalar Maxwell equations (one pair for each co-ordinate).

Obviously, there is a possibility of two directions of rotations of each quark. Then there must exist 6+6 scalar equations for proton description and also 6+6 equations for the antiproton description.

Let's find these equations without interaction, i.e. putting the interaction (mass) terms equal to zero. Using the wave function form (3.8) and the three-knot matrices (3.7) from the equation (3.5) (or (3.6)) we obtain:

$$\frac{1}{c}\frac{\partial}{\partial t}\begin{pmatrix} E_x \\ E_y \\ E_z \\ iH_x \\ iH_y \\ iH_z \end{pmatrix} + \left[\frac{\partial}{\partial x}\begin{pmatrix} 0 \\ H_z \\ -H_y \\ 0 \\ -iE_z \\ iE_y \end{pmatrix} + \frac{\partial}{\partial y}\begin{pmatrix} -H_z \\ 0 \\ H_x \\ iE_z \\ 0 \\ -iE_x \end{pmatrix} + \frac{\partial}{\partial z}\begin{pmatrix} H_y \\ -H_x \\ 0 \\ -iE_y \\ iE_x \\ 0 \end{pmatrix}\right] = 0, \qquad (3.13)$$

or

$$\begin{cases} \frac{1}{c}\frac{\partial E_x}{\partial t} - \left(\frac{\partial H_z}{\partial y} - \frac{\partial H_y}{\partial z}\right) = 0 \\ \frac{1}{c}\frac{\partial E_y}{\partial t} + \left(\frac{\partial H_z}{\partial x} - \frac{\partial H_x}{\partial z}\right) = 0 \\ \frac{1}{c}\frac{\partial E_z}{\partial t} - \left(\frac{\partial H_y}{\partial x} - \frac{\partial H_x}{\partial y}\right) = 0 \\ \frac{1}{c}\frac{\partial H_x}{\partial t} + \left(\frac{\partial E_z}{\partial y} - \frac{\partial E_y}{\partial z}\right) = 0 \\ \frac{1}{c}\frac{\partial H_y}{\partial t} - \left(\frac{\partial E_z}{\partial x} - \frac{\partial E_x}{\partial z}\right) = 0 \\ \frac{1}{c}\frac{\partial H_z}{\partial t} + \left(\frac{\partial E_y}{\partial x} - \frac{\partial E_x}{\partial y}\right) = 0 \end{cases}, \qquad (3.14)$$

As we see, the equations (3.13) and (3.14) coincide entirely with the electromagnetic wave equation system.

Let’s consider now the hypothetical three-knot equation of the three-quark particle. If the equations (3.13) describe the knots with the *x,y,z* directions, we can obtain the following equations:

$$\begin{cases} \frac{1}{c}\frac{\partial E_x}{\partial t} - \left(\frac{\partial H_z}{\partial y}\right) = 0 & a \\ \frac{1}{c}\frac{\partial H_z}{\partial t} - \left(\frac{\partial E_x}{\partial y}\right) = 0 & a' \\ \frac{1}{c}\frac{\partial E_y}{\partial t} - \left(\frac{\partial H_x}{\partial z}\right) = 0 & b \\ \frac{1}{c}\frac{\partial H_x}{\partial t} - \left(\frac{\partial E_y}{\partial z}\right) = 0 & b' \\ \frac{1}{c}\frac{\partial E_z}{\partial t} - \left(\frac{\partial H_y}{\partial x}\right) = 0 & c \\ \frac{1}{c}\frac{\partial H_y}{\partial t} - \left(\frac{\partial E_z}{\partial x}\right) = 0 & c' \end{cases} \qquad (3.15)$$

$$\begin{cases} \frac{1}{c}\frac{\partial E_z}{\partial t} + \left(\frac{\partial H_x}{\partial y}\right) = 0 & a \\ \frac{1}{c}\frac{\partial H_x}{\partial t} + \left(\frac{\partial E_z}{\partial y}\right) = 0 & a' \\ \frac{1}{c}\frac{\partial E_x}{\partial t} + \left(\frac{\partial H_y}{\partial z}\right) = 0 & b \\ \frac{1}{c}\frac{\partial H_y}{\partial t} + \left(\frac{\partial E_x}{\partial z}\right) = 0 & b' \\ \frac{1}{c}\frac{\partial E_y}{\partial t} + \left(\frac{\partial H_z}{\partial x}\right) = 0 & c \\ \frac{1}{c}\frac{\partial H_z}{\partial t} + \left(\frac{\partial E_y}{\partial x}\right) = 0 & c' \end{cases} \qquad (3.15')$$

As it is not difficult to see, each pair of equations *a,b,c* describe a separate knot; the knots of equations (3.15) are rolled up in the plains *XOZ*, *ZOY*, *YOX*, and the knots of the equations (3.15’) are rolled in the plains *XOY*, *YOZ* , *ZOX*.

## 3.4. Interaction appearance

The interaction term in the electron equation was obtained as a consequence of the electromagnetic field motion along the curvilinear trajectory (see also [1]). This coincides with the conclusions of modern physics.

The modern particle theory is also known as the gauge field theory. Without a detail discussion *we will only underline the basic property of this theory: the interactions between the particles are introduced in the field equation via the gauge transformations and this procedure is equivalent to the field vector transformations in the curvilinear space* [4,12]. These transformations lead to the covariant derivative appearance [4,12]:

$$D_\mu \psi = \left( \frac{\partial}{\partial u^\mu} - \frac{1}{2} B_\mu^{kl} M_{kl} \right) \psi , \tag{3.16}$$

which contain the interaction fields $\frac{1}{2} B_\mu^{kl} M_{kl}$, while the usual derivative these fields don't have.

**Our theory shows that the appearance of the interaction terms is not a mathematical transformation, but is bounded with the vector motion along the curvilinear trajectory.**

Actually, in the linear photon equation the interaction terms don't exist. Therefore, inside of the particle the interaction among the particle parts appears only in the instant when the photon begins to roll up. The electromagnetic form of the pair production theory of electron-positron, stated above, shows that thanks to the particle trajectory distortion, in the equation the additional terms appear.

The derivative additional term appearance follows from the general theory of the vector motion along the curvilinear trajectory. This theme was studied in the vector analyse, in the differential geometry and in the hypercomplex number theory hundred years ago [13,14] and it is well known. It must be noted that in physics there is no need in using all the mathematical results, but only the measured physical value (e.g., three co-ordinates of the space; time; electromagnetic field vectors; energy, momentum and momentum of rotation, etc.). Below we consider some conclusions of these theories.

Any vector $\vec{F}(\vec{r})$ can have the following forms [13]:

$$\vec{F}(\vec{r}) = \vec{F}(x^1, x^2, x^3) = F^1 \vec{e}_1 + F^2 \vec{e}_2 + F^3 \vec{e}_3 = F_1 \vec{e}^{\,1} + F_2 \vec{e}^{\,2} + F_3 \vec{e}^{\,3} , \tag{3.17}$$

where $F^1, F^2, F^3, F_1, F_2, F_3$ are the invariant and co-variant vector modulus and $\vec{e}^{\,i}$ *and* $\vec{e}_i$ are the basis vectors, which in general case are changed from point to point. When vector moves along the curvilinear trajectory the partial derivatives get the view:

$$\frac{\partial \vec{F}}{\partial x^j} = \frac{\partial F^i}{\partial x^j} \vec{e}_i + F^i \frac{\partial \vec{e}_i}{\partial x^j} = \frac{\partial F_i}{\partial x^j} \vec{e}^{\,i} + F_i \frac{\partial \vec{e}^{\,i}}{\partial x^j} , \tag{3.18}$$

where the following notations are used:

$$\frac{\partial \vec{e}_i}{\partial x^j} = \Gamma^k_{ij}\vec{e}_k = -\Gamma^i_{kj}\vec{e}^{\,k}, \qquad (3.19)$$

The coefficients $\Gamma^k_{ij}$ are named Christoffel symbols or bound coefficients. Thus, for the $y$- direction photon

$$\begin{cases} \vec{E} = E_3\vec{e}^{\,3} \\ \vec{H} = H_1\vec{e}^{\,1} \end{cases}, \qquad (3.20)$$

we obtain:

$$\begin{cases} \frac{1}{c}\frac{\partial \vec{E}}{\partial t} = \frac{\partial E_3}{\partial x^0}\vec{e}^{\,3} + E_3\Gamma^3_{k0}\vec{e}^{\,k} \\ \frac{1}{c}\frac{\partial \vec{H}}{\partial t} = \frac{\partial H_1}{\partial x^0}\vec{e}^{\,1} + H_1\Gamma^1_{k0}\vec{e}^{\,k} \\ \frac{\partial \vec{E}}{\partial y} = \frac{\partial E_3}{\partial x^2}\vec{e}^{\,3} + E_3\Gamma^3_{k2}\vec{e}^{\,k} \\ \frac{\partial \vec{H}}{\partial y} = \frac{\partial H_1}{\partial x^2}\vec{e}^{\,1} + H_1\Gamma^1_{k2}\vec{e}^{\,k} \end{cases}, \qquad (3.21)$$

The same we can obtain for the other directions of the photons.

*As we see here the additional terms, which the initial linear equations didn't have, have appeared. Thus, in the general case, when the electromagnetic field vectors of three-knot particles move along the curvilinear trajectories, the additional terms of the same type, which we obtained in the case of Dirac's equation, appear. Thus, we can obtain equations similar to the Standard Model Theory equations.*

As result we obtain the following equation system:

$$\begin{cases} \frac{1}{c}\frac{\partial E_x}{\partial t} - \left(\frac{\partial H_z}{\partial y} - \frac{\partial H_y}{\partial z}\right) = -i\frac{\omega_1}{c}E_x \\ \frac{1}{c}\frac{\partial E_y}{\partial t} + \left(\frac{\partial H_z}{\partial x} - \frac{\partial H_x}{\partial z}\right) = -i\frac{\omega_1}{c}E_y \\ \frac{1}{c}\frac{\partial E_z}{\partial t} - \left(\frac{\partial H_y}{\partial x} - \frac{\partial H_x}{\partial y}\right) = -i\frac{\omega_2}{c}E_z \\ \frac{1}{c}\frac{\partial H_x}{\partial t} + \left(\frac{\partial E_z}{\partial y} - \frac{\partial E_y}{\partial z}\right) = i\frac{\omega_2}{c}H_x \\ \frac{1}{c}\frac{\partial H_y}{\partial t} - \left(\frac{\partial E_z}{\partial x} - \frac{\partial E_x}{\partial z}\right) = i\frac{\omega_3}{c}H_y \\ \frac{1}{c}\frac{\partial H_z}{\partial t} + \left(\frac{\partial E_y}{\partial x} - \frac{\partial E_x}{\partial y}\right) = i\frac{\omega_3}{c}H_z \end{cases}, \qquad (3.22)$$

where $\omega_1, \omega_2, \omega_3$ are the frequencies of each knot. These values don't entirely define the charges of quarks, as in the case of electron, since the current values relate also to the rotation and twisting of field vectors.

*Note: the Christoffel symbols are not the mathematical values, but the physical values; namely, they are the currents, which appeared thanks to the gyration and twisting of the electromagnetic vectors.*

*Note also, that these additional terms play a different role in the physics. For example, in mechanics the additional term is a mass term; in electrodynamics it is an electrical current; inside of the particle it is the inner current, i.e. longitudinal non-linear fields; etc.*

## 4. Electromagnetic "two quark" equations

Obviously, the mesons must contain two knots, i.e. they must be described by "two engaged quark" equations. In the wave or oscillation physics two knot figures appear only by summing two mutually perpendicular oscillations (Fig.3) where the electric field vector $\vec{E}$ is drawn as strokes. As we see the two-knot figure has one positive and one negative knot, that is in accordance with the quark meson model. Note that this meson model's spin must be equal to zero.

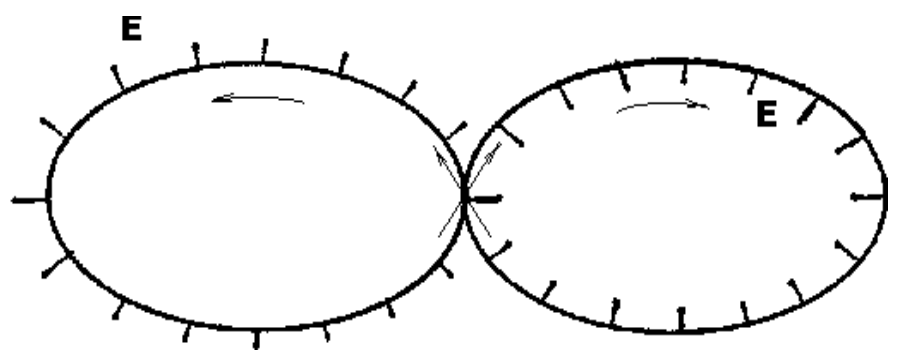

Fig.3

From above we can suppose that:

1) the meson equations must be the equation system which consists of one electron-like and one positron-like equation; and
2) these equations must be the result of the rolling up of two initial photons with mutual perpendicular directions,
3) the electric vectors of both photons must lie in one plain, e.g., of one photon with $y$-direction, which contains $E_x, H_z$, and other photon with $x$-direction, which contains $E_y, H_z$ field vectors (see Fig.4).

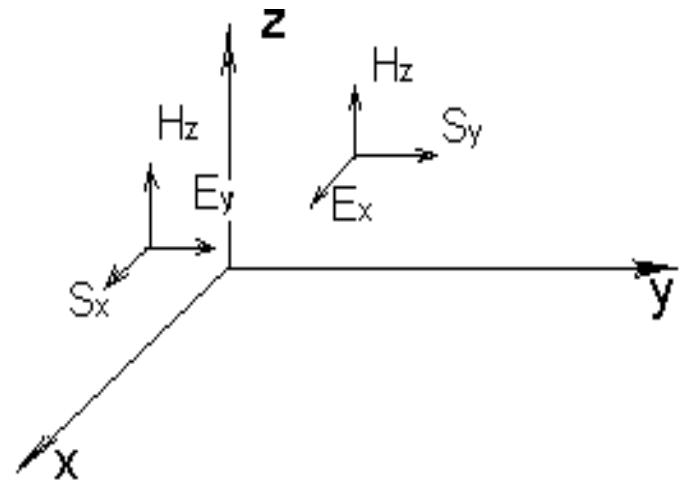

Fig.4

Obviously there are a lot of building variants of such pairs. The simplest way for the two-knot (meson) equation construction is to use the electron-positron equation or the three-knot (baryon) equations. For example, for the building of a two-knot particle in the $XOY$ plain it is enough to use the above field vectors, making equal to zero the field vectors of photon with the Poynting vector having $z$-direction.

## 5. About symmetry breaking and mass appearance

Among the QCD and electroweak theory there is one serious difference: the interaction particles of QCD (gluons) are massless, and the interaction particles of electroweak theory (intermediate bosons) have big masses. The electroweak theory is built in the same way as the QCD, and until one point the electroweak interaction particles don’t have masses. For the masses acquirement in the electroweak Lagrangian one function is introduced - a single complex Higgs doublet, which describes the Higgs boson – a hypothesised spinless particle **H**. Then, some mathematical transformation, which is named the spontaneous symmetry breaking, is realised. Thanks to the Higgs boson vacuum interaction, the massless interaction particles acquire the masses, but the Higgs boson doesn’t change; thus the Higgs boson acts as the reaction catalyst.

The problem exists: is there a particle in our photon symmetry breaking theory, whose action is similar to the Higgs boson action?

Let’s show that such particle really exists, although it has other properties from the Higgs boson. Consider newly the pair production reaction (2.1) in form:

$$\gamma + N \rightarrow N + e^{+} + e^{-}, \qquad (4.1)$$

where $N$ is the nucleus of atom (e.g., proton) or other charged particle. For the Lagrangian of this reaction we can write conventionally:

$$L(\gamma + N) = L(N + e^{+} + e^{-}), \qquad (4.2)$$

or

$$L(\gamma) + L(N) = L(N) + L(e^{+} + e^{-}), \qquad (4.3)$$

Of course, the nucleus is not the Higgs boson, but it plays the same role as a catalyst in the particle transformations.

Note also that the relation (4.1) can be considered as a description of the wave refraction, where the term $L(N)$ corresponds to the medium with some refraction index and can be described as the dispersion matrix for the transformation from the linear wave to the gyration wave.

# Appendix. Hadron models

We have considered above the hadrons as particles, consisting of two or three knots. The equation of one knot is the Dirac equation that has a harmonic solution. Therefore, it can be supposed that the hadrons are the superposition of two or three harmonic oscillations. On other words, the hadrons are the space wave packets. According to the Schreudinger [15] (see also [16], section 6.1) the wave packets, built from harmonic waves (oscilations), don’t have a dispersion, i.e. they are stable. Thus, we can, as a fist approximation, build the hadrons model as the space packet of the harmonics superposition.

Of course, the below models differ a lot from the real models and can not be used for calculation of the particle features. But they give some representation about them.

The models were constructed using Mathcad-program.

## A1. “Three quarks” model

Thus, we suppose that the three-knot model is built from three harmonics oscillation. Let’s choose the following oscillation parameters:

$$\omega_1 = 3, \quad \omega_2 = 2, \quad \omega_3 = 3 \qquad N := 200$$

$$\phi_1 = \frac{\pi}{2}, \quad \phi_2 = \frac{\pi}{2}, \quad \phi_3 = 0 \qquad j := 0..N$$

$$r_1 = 2, \quad r_2 = 2, \quad r_3 = 2 \qquad k := 0..N$$

$$t_j := j \cdot 2 \cdot \frac{\pi}{N} \qquad v_k := k$$

$$X_{k,j} := r_1 \cdot \sin(\omega_1 \cdot t_j - \phi_1)$$

$$Y_{k,j} := r_2 \cdot \sin(\omega_2 \cdot t_j - \phi_2)$$

$$Z_{k,j} := r_3 \cdot \sin(\omega_3 \cdot t_j - \phi_3)$$

We obtain the following three-knots figure A1:

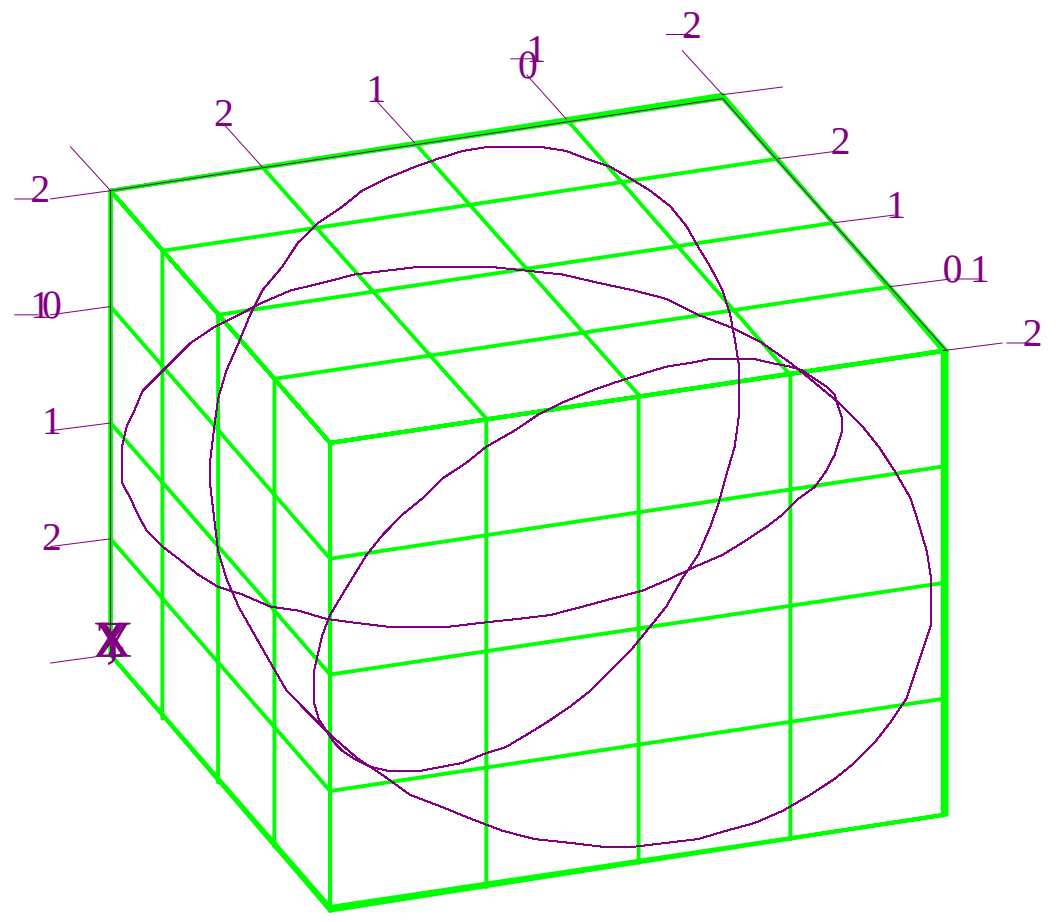

Fig. A1

To show the field plain gyration and twisting we change the parameter $t_j$ to $t_j := \frac{j}{2.2}$. Then we obtain figure A2:

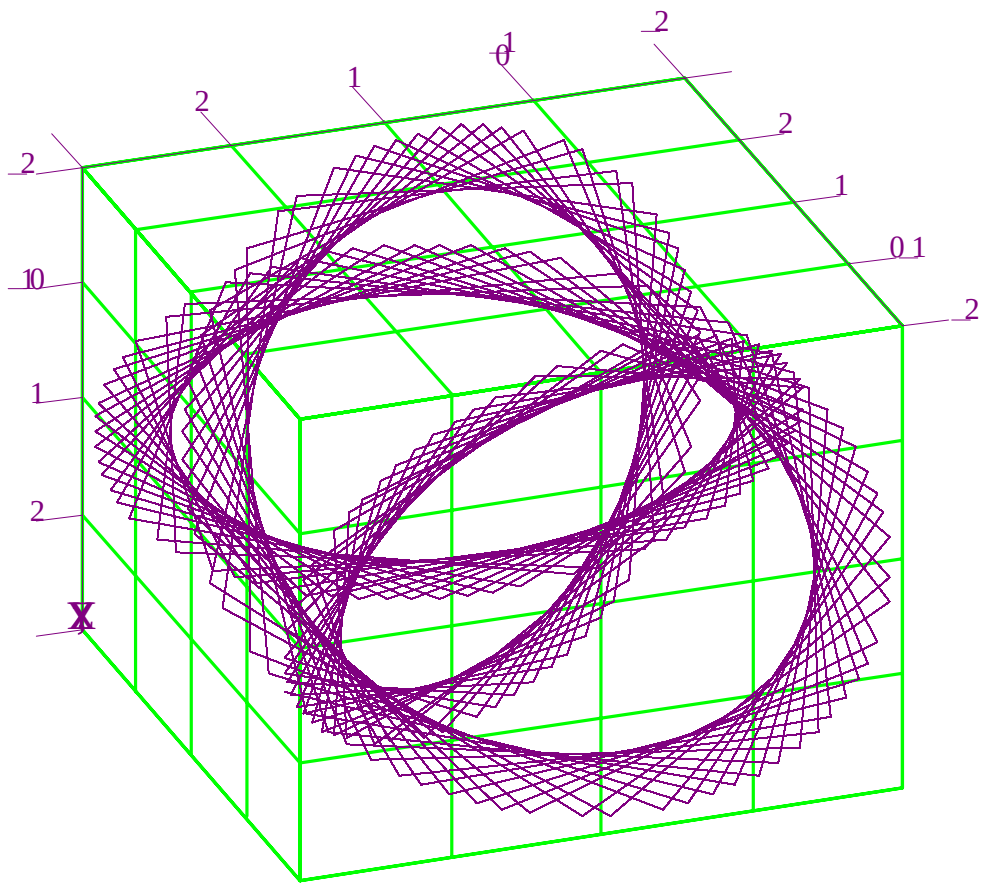

Fig. A2

**A2. “Two quarks” model**

To build the two-knot (meson) model in the above proton model equations we choose the following new parameters:

$$\omega_1 = 1 \text{ and } \phi_1 = 0,$$

and put $Z_{k,j} := 0$. Then we obtain the figure A3:

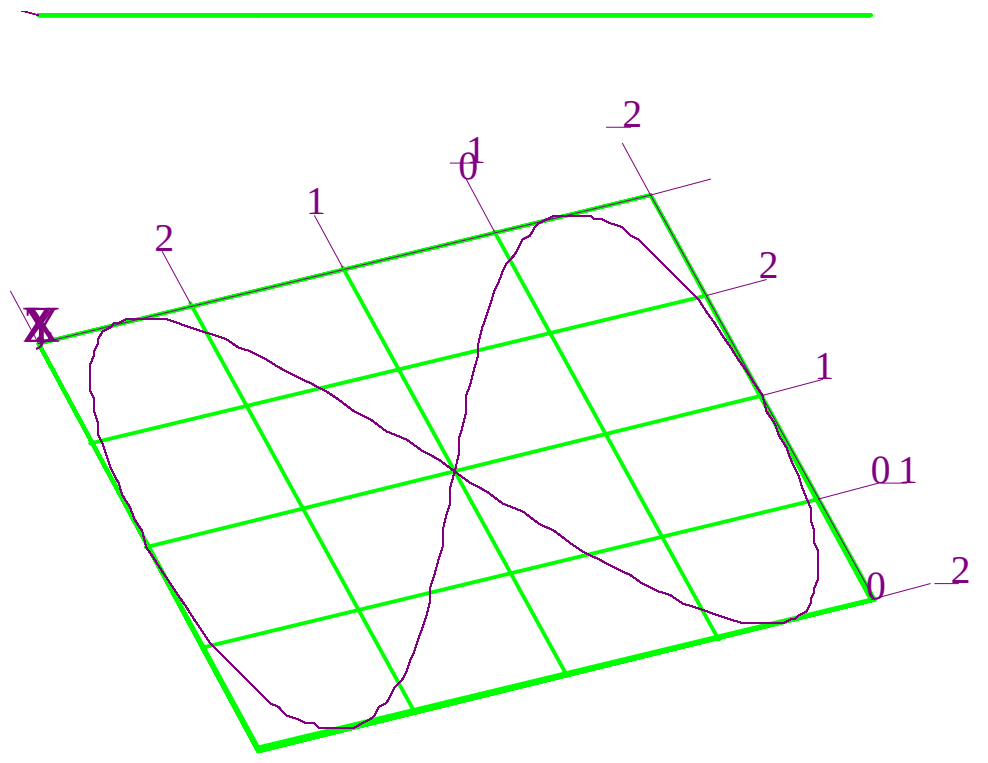

Fig.A3

We hope that the further investigations will allow us to build real models, which will give us the opportunity to calculate the particle features.

## Conclusion

Our scope to show that Maxwell's equations can describe the composite space figures, which consist of two or three knots and have the hadrons properties, is fulfilled.

Of course we can not yet answer all questions about hadrons, but if the above theory is right we can understand better what are the quark and gluon, why free quarks and gluons do not exist, why quarks have fractional charges, why gluons interact with each other and can produce the particles; and many others.

## References

1. Kyriakos, A.G.(http://arXiv.org/abs/quant-ph/0204037), 2002.
2. Kyriakos, A.G.(http://arXiv.org/abs/quant-ph/0204134), 2002
3. Kyriakos, A.G.(http://arXiv.org/abs/quant-ph/0205075), 2002.
4. Lewis H. Ryder. *Quantum field theory*. Cambridge university press, 1985.
5. L.B.Okun. Elementary particle physics (in Russian). Moscow, 1988.
6. A.Pich. Aspects of quantum chromodynamics. (arXiv: hep-ph/0001118 v.1, 13 Jan.2000)
7. L.T. Schiff. Quantum Mechanics. 2nd edition, McGraw-Hill Book Company, Jnc, New York, 1955.
8. A.Sokoloff, D.Iwanenko. Quantum field theory (in Russian). Moskwa-Leningrad, 1952.
9. M.-A. Tonnelat. Les Principes de la Theorie Electromagnetique et de la Relativite. Masson et C. Editeurs, Paris, 1959.
10. W.J.Archibald.Canad. Journ.Phys., **33**, 565 (1955)
11. A.I. Achieser, W.B. Berestetski. Quantum electrodynamics. Moscow,1969.
12. F.A.Kaempffer. Concepts in quantum mechanics. Academic press, N.Y and London, 1965.
13. G.A. Korn, Th.M. Korn. Mathematical handbook for scientists and engineers. McGraw-Hill Book Co., 1961.
14. E. Madelung. Die mathematischen hilfsmittel des physikers. Springer verlag, Berlin, 1957
15. E.Schreudinger. Naturwissenschaften, Bd.14, S.664-666 (1926)
16. M.Jammer. The conceptualdevelopment of quantum mechanics. McGraw-Hill book Company, 1967.